\documentclass[journal]{IEEEtran}
\usepackage{amsmath}
\usepackage{amssymb}
\usepackage{amsthm}
\usepackage{algorithm}
\usepackage{etoolbox}
\usepackage{stfloats,hyperref}
\usepackage{algpseudocode}
\usepackage{algorithm}
\usepackage{algpseudocode}
\algnewcommand\Input{\item[\textbf{Input:}]}
\algnewcommand\Output{\item[\textbf{Output:}]}
\usepackage{bm}
\usepackage{graphicx,color}
\usepackage{cite}
\usepackage{booktabs}
\IEEEoverridecommandlockouts
\usepackage{graphicx}    
\usepackage{subcaption}  
\usepackage[font=scriptsize,labelfont=bf]{caption}

 \usepackage[dvipsnames]{xcolor}
 \usepackage{textcomp}
\usepackage{tikz}

\definecolor{lime}{HTML}{A6CE39}
\DeclareRobustCommand{\orcidicon}{%
    \begin{tikzpicture}
    \draw[lime, fill=lime] (0,0)
    circle [radius=0.16]
    node[white] {{\fontfamily{qag}\selectfont \tiny ID}};
    \draw[white, fill=white] (-0.0625,0.095)
    circle [radius=0.007];
    \end{tikzpicture}
    \hspace{-2mm}
}

\foreach \x in {A, ..., Z}{%
    \expandafter\xdef\csname orcid\x\endcsname{\noexpand\href{https://orcid.org/\csname orcidauthor\x\endcsname}{\noexpand\orcidicon}}
}

\hypersetup{
     colorlinks = true,
     linkcolor = blue,
     anchorcolor = blue,
     citecolor = blue,
     filecolor = blue,
     urlcolor = blue
     }

\title{\huge Robust Quantum-MUSIC for DoA Estimation \\ Using Rydberg Atomic Receiver Arrays}

\author{Sourav Banerjee\orcidB{}, Neel Kanth Kundu\orcidA{},~\IEEEmembership{Member,~IEEE}, and Prajwalita Borah
\thanks{The work of Neel Kanth Kundu was supported in part by the DST INSPIRE Faculty Fellowship (Reg. No.: IFA22-ENG 344), ANRF PM Early Career Research Grant (ANRF/ECRG/2024/000324/ENS), and IIT Delhi's New Faculty Seed Grant (MI02855G). \textit{(Corresponding author: Neel Kanth Kundu.)}
}
\thanks{Sourav Banerjee and Prajwalita Borah are with the Center for Applied Research in Electronics (CARE), Indian Institute of Technology (IIT) Delhi, New Delhi-110016, India (e-mail: crz248112@care.iitd.ac.in, crz248470@care.iitd.ac.in).

Neel Kanth Kundu is with CARE and Bharti School of Telecommunication Technology and Management, IIT Delhi, New Delhi, India. He is also an honorary fellow at the Department of Electrical and Electronic Engineering, University of Melbourne, Melbourne, VIC-3010, Australia (e-mail: neelkanth@iitd.ac.in). }
}

\makeatletter
\patchcmd{\@maketitle}
  {\addvspace{0.5\baselineskip}\egroup}
  {\addvspace{-1\baselineskip}\egroup}
  {}
  {}
\makeatother

\begin{document}
\maketitle
\begin{abstract}
Quantum wireless sensing using Rydberg atomic receivers enables
high-sensitivity signal acquisition direction-of-arrival
(DoA) estimation. However, it suffers from a fundamental
limitation, where only the magnitude of the received signal is observable.
The recently proposed Quantum-MUSIC algorithm addresses this problem by
recovering phase information through alternating minimization and
subsequently applying the MUSIC algorithm for 
DoA estimation.
However, the existing approach relies on an $\ell_2$-norm phase
retrieval step, making it highly sensitive to outlier measurements
produced by hardware faults, sensor saturation, or adversarial
interference.
In this letter, we propose a \emph{Robust Quantum-MUSIC}
(RobQMUSIC) framework that replaces the $\ell_2$-norm with an
$\ell_1$-norm formulation.
The resulting weighted phase-retrieval problem is solved efficiently
via an Iteratively Reweighted Least Squares (IRLS) scheme embedded
within the alternating minimization loop, requiring no increase in
structural complexity relative to the baseline algorithm.
Simulation results demonstrate that RobQMUSIC achieves near-identical
DoA estimation accuracy to Quantum-MUSIC under ideal conditions, while
maintaining robust performance over a wide range of outlier
contamination levels at which Quantum-MUSIC fails entirely.
\end{abstract}


\section{Introduction}
\label{sec:intro}

Recent years have witnessed a paradigm shift from conventional
wireless sensing toward quantum-enabled sensing, driven by the need
for ultra-high sensitivity and precision in next-generation
communication and radar systems~\cite{chen2026harnessingrydbergatomicreceivers,banerjee2026}.
In particular, atomic RF receivers based on highly excited Rydberg
states have emerged as a promising technology for quantum wireless
sensing, owing to their ability to interact directly with
electromagnetic fields over a broad frequency
spectrum~\cite{anderson2021,fancher2021}.
Unlike classical antennas, these quantum RF sensors exploit the atom-field
interactions to achieve exceptionally high sensitivity, thereby opening new
avenues for high-resolution channel estimation and direction of arrival (DoA)
sensing~\cite{robinson2021}.

Despite growing interest, the integration of atomic receivers into practical wireless sensing systems remains at an early stage.
Existing works have primarily addressed simplified scenarios such as single-target sensing via geometric phase
relationships~\cite{robinson2021}.
While such approaches demonstrate feasibility, they fail to
generalize to realistic multi-user environments.
Although several experimental studies have validated quantum wireless
sensing~\cite{mao2023}, accurate multi-user DoA estimation remains
challenging owing to the magnitude-only measurement constraint
of the Rydberg atomic RF receiver.

To address this challenge, Quantum-MUSIC was recently proposed as a
signal-processing framework for multi-user quantum wireless sensing
with atomic RF receivers~\cite{kim2025}.
The algorithm first recovers the channel matrix from
magnitude-only measurements via a modified, biased
Gerchberg-Saxton (GS) phase-retrieval procedure~\cite{cui2025},
followed by the application of the classical MUSIC algorithm~\cite{schmidt1986} for
multi-user DoA estimation.
Numerical results in~\cite{kim2025} demonstrated that Quantum-MUSIC
outperforms conventional RF-domain MUSIC, establishing a
foundational framework for Rydberg-atom-based DoA estimation problems.

A critical limitation of Quantum-MUSIC, however, is its sensitivity
to measurement imperfections.
In practice, atomic receiver outputs are corrupted by hardware
malfunctions, sensor saturation, intermittent atomic-excitation
failures, or adversarial interference, resulting in large-amplitude
additive outliers in a sparse subset of measurements.
Since the channel-recovery step of the Quantum-MUSIC algorithm minimizes an
$\ell_2$-norm residual—which penalizes errors quadratically—even a
small fraction of such outliers can dominate the cost and severely
degrade the recovered channel matrix and the ensuing DoA estimates.

Robust estimation via $\ell_1$-norm minimization is a well-established
remedy for outlier sensitivity~\cite{candes2008introduction,boyd2004convex}.
Linear penalization of residuals limits the influence of large
outliers without sacrificing accuracy on inlier measurements.
Although $\ell_1$ problems are not directly solvable in closed form, the Iteratively Reweighted Least Squares (IRLS) algorithm provides a computationally tractable approximation by
iteratively solving a sequence of diagonal-reweighted
$\ell_2$ problems~\cite{daubechies2010,lai2013}.

In this letter, we propose a robust extension of
Quantum-MUSIC (RobQMUSIC) that replaces the $\ell_2$ phase-retrieval step with
an $\ell_1$-IRLS formulation.
The proposed method retains the same spectral-method initialization
and alternating-minimization structure as~\cite{kim2025}, adding only
an inner IRLS loop that reweights each snapshot according to its
current residual magnitude.
This modification is simple to implement, adds negligible
computational overhead per outer iteration, and confers strong
robustness to sparse large-amplitude outliers.

\emph{Notation:}
Boldface lowercase and uppercase letters denote vectors and matrices,
respectively.
$(\cdot)^H$ and $(\cdot)^T$ denote conjugate transpose and transpose.
$|\cdot|$ applied element-wise denotes the modulus; $\|\cdot\|_1$
and $\|\cdot\|_2$ denote the $\ell_1$- and $\ell_2$-norms.
$\odot$ and $\circ$ denote the Hadamard (element-wise) product.
$\mathcal{CN}(\boldsymbol{\mu},\sigma^2\mathbf{I})$ denotes a
circularly-symmetric complex Gaussian distribution.

\section{Quantum Physics of the Rydberg Atomic Receiver}
\label{sec:quantum}


For completeness, we briefly review the relevant Rydberg atom physics principles 
from~\cite{kim2025}. In the considered system, $K$ single-antenna users transmit RF
signals to an atomic receiver array comprising $M$ vapor cells,
each filled with Rydberg atoms.
Unlike conventional receivers, information is not processed in the
electrical domain; instead it is encoded in the quantum-state
transitions of the atoms.
The total EM field incident on the $m$-th vapor cell is formed by
the superposition of all users' signals:
\begin{equation}
  E_m(t)
  = \sum_{k=1}^{K}
    \boldsymbol{\epsilon}\sqrt{P_k}\,\rho_{k}\,s_k
    \cos\!\bigl(\omega t+\varphi_{m,k}\bigr),
  \label{eq:em_field}
\end{equation}
where $\boldsymbol{\epsilon}$, $\rho_{k}$, $\varphi_{m,k}$,
$P_k$, and $s_k$ denote the polarization direction, path loss,
phase shift, transmit power, and transmitted signal of the $k$-th
user at the $m$-th cell, respectively. This field drives quantum-state transitions whose strength is
characterised by the \emph{Rabi frequency}, observable via
electromagnetically induced transparency (EIT) and Autler--Townes
(AT) splitting~\cite{liu2023}.
The Rabi frequency at the $m$-th cell is
\begin{equation}
  \Omega_m
  = \left|\sum_{k=1}^{K}
      \frac{1}{\hbar}\,\boldsymbol{\mu}_{eg}^{H}
      \boldsymbol{\epsilon}
      \sqrt{P_k}\,\rho_{k}\,s_k\,e^{-j\varphi_{m,k}}
    \right|,
  \label{eq:rabi}
\end{equation}
where $\hbar$ is the reduced Planck constant and $\boldsymbol{\mu}_{eg}$
is the transition dipole moment.
Equation~\eqref{eq:rabi} shows that the Rabi frequency inherently
encodes the full channel---power, path loss, phase shift, and
polarization---transforming the classical RF reception problem into
a quantum measurement problem whose output is a \emph{magnitude-only}
observable.

\section{System Model}
\label{sec:system}

\subsection{Signal and Channel Model}

\begin{figure}
    \centering
    \includegraphics[width=0.8\linewidth]{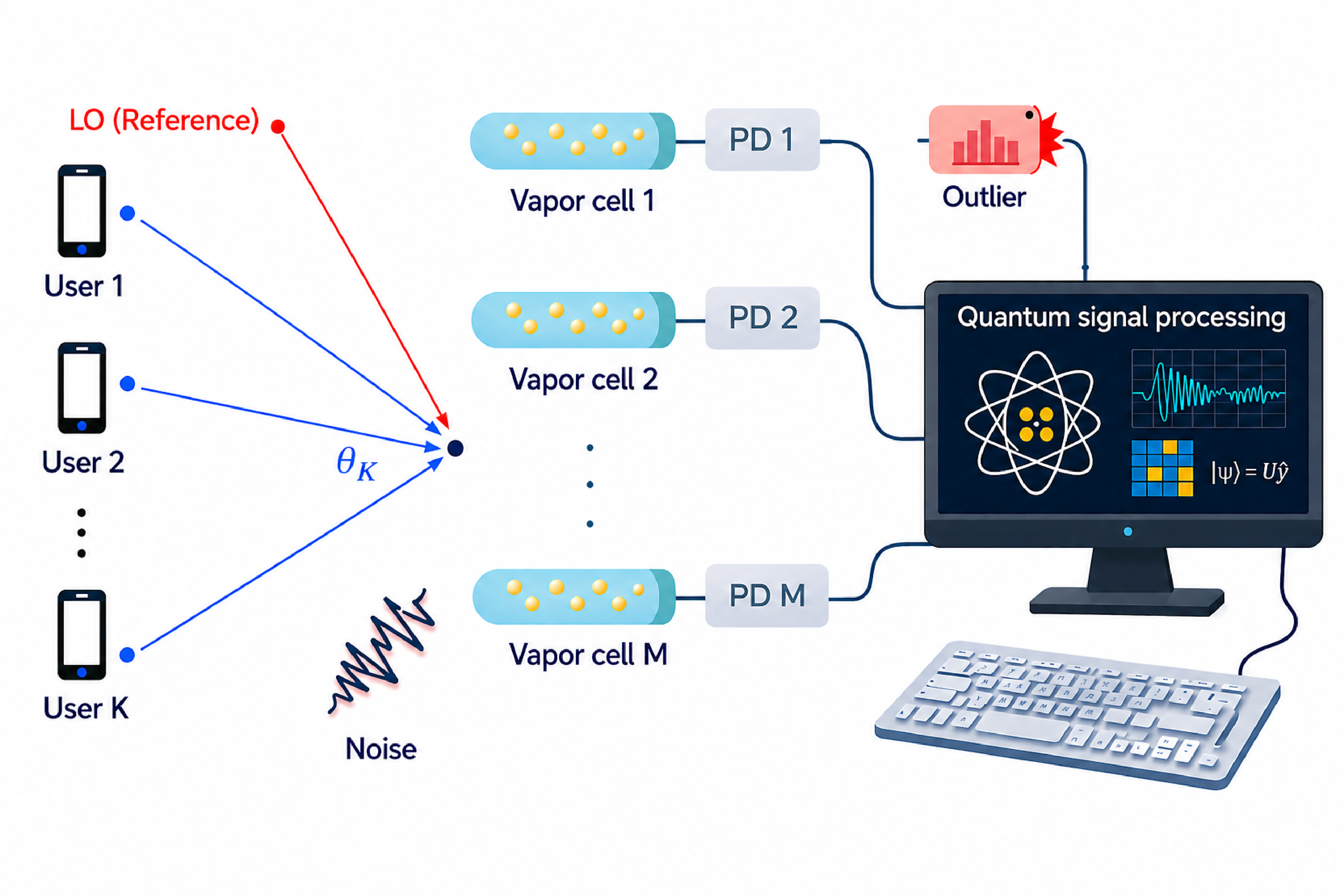}
    \caption{Schematic of the multi-user DoA estimation system using Rydberg atomic receiver array in the presence of outliers.}
    \label{fig:system_model}
\end{figure}
As shown in Fig.~\ref{fig:system_model}, $K$ users transmit signals toward a uniform linear array (ULA) made of Rydberg receivers. A local oscillator (LO) is also present as a reference signal. These RF signals are converted into optical responses via the interaction of electromagnetic fields with Rydberg atoms in vapor cells. The optical signals are detected by photodetectors (PDs), producing electrical measurement signals.

Let $\mathbf{s}\in\mathbb{C}^K$ be the transmitted signal vector.
The complex baseband signal at the $m$-th sensor is
\begin{equation}
  y_m = \mathbf{s}^{H}\mathbf{h}_m + b_m + n_m,
  \label{eq:rx_signal}
\end{equation}
where $\mathbf{h}_m\in\mathbb{C}^K$ is the channel vector,
$b_m = \frac{1}{\hbar}\boldsymbol{\mu_{eg}}^H \boldsymbol{\epsilon}_{LO} \sqrt {P_b}\in\mathbb{C}$ is a known deterministic bias from the
LO reference, and
$n_m\sim\mathcal{CN}(0,\sigma_n^2)$ is the quantum shot noise (QSN). Here, the relation between the Rabi frequency and complex baseband signal is $\Omega_m = \lvert y_m \rvert$.

\subsection{Steering-Vector Parametrization}

Assuming a narrowband signal in the far-field propagation and a ULA with half-wavelength spacing
$d=\lambda/2$, the channel vector of the $k^{\rm th}$ user is the steering vector 
\begin{equation}
  \mathbf{a}(\theta_k)
  = \alpha\,\bigl[1,\;
    e^{-j\pi\sin\theta_k},\;\ldots,\;
    e^{-j\pi(M-1)\sin\theta_k}\bigr]^T \in\mathbb{C}^M,
  \label{eq:steering}
\end{equation}
where $\theta_k$ is the DoA of the $k$-th user and $\alpha = \frac{1}{\hbar} \boldsymbol{\mu}_{eg}^{H}\boldsymbol{\epsilon} \sqrt{P_k}\rho_k $ subsumes path loss and polarization.
Stacking all $K$ steering vectors gives the array manifold matrix
$\mathbf{A}(\boldsymbol{\theta})
 =[\mathbf{a}(\theta_1),\ldots,\mathbf{a}(\theta_K)]
 \in\mathbb{C}^{M\times K}$, which can also be expressed as follows:
 \begin{equation}
\mathbf{A}(\theta)
=
\begin{bmatrix}
\text{---}\,\mathbf{h}_1^{T}\,\text{---} \\
\text{---}\,\mathbf{h}_2^{T}\,\text{---} \\
\vdots \\
\text{---}\,\mathbf{h}_M^{T}\,\text{---}
\end{bmatrix}
\end{equation}
By performing channel estimation, the DoA vector $\boldsymbol{\theta}=[\theta_1,\ldots,\theta_K]^T$ can be further  estimated.

\subsection{Multi-Snapshot Magnitude-Only Measurement Model}

Collecting $P$ independent pilot snapshots, the pilot matrix is given by
$\mathbf{S}=[\mathbf{s}_1,\ldots,\mathbf{s}_P]\in\mathbb{C}^{K\times P}$.
Stacking all sensor outputs yields the \emph{measurement matrix}
\begin{equation}
  \mathbf{Z}
  = \bigl[\mathbf{Z}_{:,1},\ldots,\mathbf{Z}_{:,P}\bigr]
  \in\mathbb{R}^{M\times P}_+,
  \quad
  \mathbf{Z}_{:,p} = |\mathbf{A}(\boldsymbol{\theta})\mathbf{s}_p
                   +\mathbf{b}+\mathbf{n}_p|,
  \label{eq:meas_matrix}
\end{equation}
where the modulus is applied element-wise.
Equivalently, $\mathbf{z}_m^T$ is the $m$-th row that satisfies $\mathbf{z}_m = |\mathbf{S}^H\mathbf{h}_m+\mathbf{b}_m+\mathbf{n}_m|$,
where $\mathbf{b}_m = b_m\mathbf{1}_P$.

\subsection{Corrupted Measurement Model}

In practice, a sparse subset of entries of $\mathbf{Z}$ is corrupted
due to hardware faults, saturation, or interference.
Let $\mathcal{S}\subset\{1,\ldots,M\}\times\{1,\ldots,P\}$ be the
unknown support set with $|\mathcal{S}|=S\ll MP$.
The observed matrix is
\begin{equation}
  \widetilde{\mathbf{Z}} = \mathbf{Z} + \mathbf{E},
  \label{eq:corrupted_meas}
\end{equation}
where $\mathbf{E}\in\mathbb{R}^{M\times P}$ is a sparse matrix with
$E_{m,p}\neq 0$ iff $(m,p)\in\mathcal{S}$.
The non-zero entries represent \emph{outliers} of arbitrary magnitude
and sign with no assumed distribution.
Because $\ell_2$-based channel recovery is sensitive to such
large-magnitude corruption, we formulate the channel estimation
problem as an $\ell_1$-norm minimization:
\begin{equation}
  \min_{\mathbf{h}_m}
  \bigl\|\tilde{\mathbf{z}}_m
        -\bigl|\mathbf{S}^H\mathbf{h}_m+\mathbf{b}_m\bigr|\bigr\|_1,
  \quad m=1,\ldots,M,
  \label{eq:main_opt}
\end{equation}
where $\tilde{\mathbf{z}}_m^T$ denotes the $m$-th row of
$\widetilde{\mathbf{Z}}$.

\section{Proposed Robust Quantum-MUSIC Algorithm}
\label{sec:algorithm}

\subsection{Alternating Minimization Formulation}

Problem~\eqref{eq:main_opt} is non-convex due to the modulus
operator.
Following~\cite{kim2025}, we introduce an auxiliary phase vector
$\angle\mathbf{u}_m$ and reformulate the problem as the alternating
minimization
\begin{equation}
  \min_{\mathbf{h}_m,\,\angle\mathbf{u}_m}
  \bigl\|\tilde{\mathbf{z}}_m\circ e^{j\angle\mathbf{u}_m}
        -\mathbf{S}^H\mathbf{h}_m-\mathbf{b}_m\bigr\|_1,
  \label{eq:l1_alt}
\end{equation}
which decouples into a \emph{phase update} and an
\emph{amplitude update} step, each convex when the other is fixed.

\subsection{Spectral Initialisation}

To avoid convergence to poor local minima, the channel estimate
is initialised using the spectral method of~\cite{candes2015}.
Define the expanded signal matrix
$\bar{\mathbf{S}}=[\mathbf{S}^H,\mathbf{b}_m]^H\in\mathbb{C}^{(K+1)\times P}$
and the expanded channel vector
$\bar{\mathbf{h}}_m=[\mathbf{h}_m^H,1]^H\in\mathbb{C}^{K+1}$.
The expanded covariance matrix is
\begin{equation}
  \bar{\mathbf{R}}
  = \sum_{p=1}^{P}\bar{z}_{m,p}\,
    \bar{\mathbf{s}}_p\bar{\mathbf{s}}_p^H
  \in\mathbb{C}^{(K+1)\times(K+1)}.
  \label{eq:cov_init}
\end{equation}
Letting $\mathbf{v}$ be the principal eigenvector of $\bar{\mathbf{R}}$,
the initial estimate is $\bar{\mathbf{a}}_m^{0}=\bar{r}\,\mathbf{v}$
with scalar
$\bar{r}=|\mathbf{v}^H\bar{\mathbf{S}}|\tilde{\mathbf{z}}_m||\,/\,
          \|\bar{\mathbf{S}}^H\mathbf{v}\|_2^2$,
and $\mathbf{h}_m^{0}=\bar{\mathbf{h}}_m^{0}(1:K)$.

\subsection{Phase Update}

For fixed $\mathbf{h}_m^{n-1}$, the phase update is obtained as
\begin{equation}
  \angle\mathbf{u}_m^n
  = \angle\!\bigl(\mathbf{S}^H\mathbf{h}_m^{n-1}+\mathbf{b}_m\bigr),
  \label{eq:phase_update}
\end{equation}
and setting $\tilde{\mathbf{z}}_m^n
= \tilde{\mathbf{z}}_m\circ e^{j\angle\mathbf{u}_m^n}$.

\subsection{Amplitude Update via IRLS}

For fixed $\tilde{\mathbf{z}}_m^n$, \eqref{eq:l1_alt} reduces to
the $\ell_1$ weighted least-squares problem
\begin{equation}
  \min_{\mathbf{h}_m}
  \bigl\|\tilde{\mathbf{z}}_m^n-\mathbf{S}^H\mathbf{h}_m-\mathbf{b}_m\bigr\|_1.
  \label{eq:l1_amp}
\end{equation}
We solve~\eqref{eq:l1_amp} via IRLS, which iterates the following
two sub-steps for $t=1,\ldots,T$:
\begin{enumerate}
  \item \textbf{Weight update:}
    Compute the residual
    $\boldsymbol{\varrho}^{(t)}
     =\tilde{\mathbf{z}}_m^n-\mathbf{S}^H\mathbf{h}_m^{n(t)}-\mathbf{b}_m$
    and set diagonal weights
    \begin{equation}
      w_p^{(t+1)}
      = \frac{1}{\bigl|\varrho_p^{(t)}\bigr|+\epsilon},
      \quad p=1,\ldots,P,
      \label{eq:irls_weights}
    \end{equation}
    where $\epsilon>0$ is a small regularization constant that
    prevents division by zero and also stabilizes convergence
    near the solution.
    Entries with large residuals—corresponding to outliers—receive
    small weights and are thus down-weighted in the subsequent step.
  \item \textbf{Weighted least-squares solution:}
    \begin{equation}
      \mathbf{h}_m^{n(t+1)}
      = \bigl(\mathbf{S}\mathbf{W}^{(t)}\mathbf{S}^H
              \bigr)^{-1}
        \mathbf{S}\mathbf{W}^{(t)}
        \bigl(\tilde{\mathbf{z}}_m^n-\mathbf{b}_m\bigr),
      \label{eq:wls}
    \end{equation}
where $\mathbf{W}^{(t)}=\mathrm{diag}(w_1^{(t)},\ldots,w_P^{(t)})$.
\end{enumerate}
After $T$ IRLS iterations, the amplitude estimate is set to
$\mathbf{h}_m^n=\mathbf{h}_m^{n(T)}$.

\subsection{DoA Estimation via MUSIC}

After $N$ outer iterations yield $\hat{\mathbf{h}}_m=\mathbf{h}_m^N$
for all $m$, the estimated channel matrix
$\hat{\mathbf{H}}=[\hat{\mathbf{h}}_1,\ldots,\hat{\mathbf{h}}_m]^H$
is used to form the sample covariance matrix
\begin{equation}
  \mathbf{R} = \frac{1}{P}\hat{\mathbf{H}}\hat{\mathbf{H}}^H
  \in\mathbb{C}^{M\times M}.
  \label{eq:cov_music}
\end{equation}
Eigenvalue decomposition of $\mathbf{R}$ yields the signal and noise subspaces $\mathbf{U}_S\in\mathbb{C}^{M\times K}$ and $\mathbf{U}_N\in\mathbb{C}^{M\times(M-K)}$, respectively.
For an ULA with $M$ sensors, inter-element spacing $d \approx \frac{\lambda}{2}$, and wavelength $\lambda$, the far-field steering vector is given by
$\mathbf{a}(\theta)
= \frac{1}{\sqrt{M}}
\begin{bmatrix}
1 & e^{j \pi \sin\theta} & e^{j 2\pi \sin\theta} & \cdots & e^{j (M-1)\pi \sin\theta}
\end{bmatrix}.$
The Quantum-MUSIC pseudo-spectrum is given by
\begin{equation}
P_Q(\theta)
= \frac{1}{\mathbf{a}^H(\theta)\mathbf{U}_N\mathbf{U}N^H\mathbf{a}(\theta)},
\label{eq:music_spectrum}
\end{equation}
and the DoA estimates $\{\hat\theta_k\}_{k=1}^K$ are identified as the $K$ largest peaks of $P_Q(\theta)$.


The complete robust quantum MUSIC procedure is summarized in Algorithm~\ref{alg:robqmusic}.

\subsection{Computational Complexity}

The computational complexity of RobQMUSIC consists of four main stages. Spectral initialization, including construction of the expanded covariance matrices and principal eigenvector extraction for all $M$ sensors, requires $\mathcal{O}(MK^{2}P)$. The phase update step costs $\mathcal{O}(KP)$ per sensor per outer iteration, leading to a total complexity of $\mathcal{O}(MNKP)$. For the IRLS amplitude update, each inner iteration is dominated by forming the weighted normal matrix and solving a $K\times K$ system, resulting in complexity $\mathcal{O}(K^{2}P+K^{3})$; with $T$ IRLS iterations over $N$ outer iterations and $M$ sensors, the total cost becomes $\mathcal{O}(MNT(K^{2}P+K^{3}))$. The final MUSIC spectral search, including covariance formation, eigendecomposition, and pseudo-spectrum evaluation over $G_{\theta}$ grid points, requires $\mathcal{O}(M^{3}+G_{\theta}M^{2})$. Therefore, the overall computational complexity of RobQMUSIC is dominated by
\begin{equation}
\mathcal{O}\!\left(
M\left(
K^{3}
+NTPK^{2}
+M^{2}
+G_{\theta}M
\right)
\right).
\end{equation}
Compared with Quantum-MUSIC~\cite{kim2025}, RobQMUSIC introduces only an additional multiplicative factor $T$ in the phase-retrieval stage. Since $T$ is a small fixed constant and typically $K \ll P$, this additional overhead is negligible while providing improved robustness.

\begin{algorithm}[t]
\caption{Robust Quantum-MUSIC (RobQMUSIC)}
\label{alg:robqmusic}
\begin{algorithmic}[1]
\Input $\mathbf{S}$, $\mathbf{b}$, $\widetilde{\mathbf{Z}}$,
       $N$, $T$, $\epsilon$
\Output $\{\hat\theta_k\}_{k=1}^K$ from $K$ largest peaks of
        $P_Q(\theta)$

\For{$m=1$ \textbf{to} $M$}
  \State Form $\bar{\mathbf{S}}=[\mathbf{S}^H,\mathbf{b}_m]^H$;
         compute $\bar{\mathbf{R}}$ via~\eqref{eq:cov_init}
  \State Obtain $\mathbf{h}_m^0$ via spectral initialisation
         (Section~IV-B)

  \For{$n=1$ \textbf{to} $N$}
    \State \textbf{Phase update:} $\tilde{\mathbf{z}}_m^n
           \leftarrow\tilde{\mathbf{z}}_m\circ
           e^{j\angle(\mathbf{S}^H\mathbf{h}_m^{n-1}+\mathbf{b}_m)}$

    \State Initialise $w_p^{(1)}=1,\; p=1,\ldots,P$

    \For{$t=1$ \textbf{to} $T$}
      \State Compute residual
             $\boldsymbol{\varrho}^{(t)}
              =\tilde{\mathbf{z}}_m^n
               -\mathbf{S}^H\mathbf{h}_m^{n(t)}-\mathbf{b}_m$

      \State Update weights via~\eqref{eq:irls_weights}

      \State Solve WLS via~\eqref{eq:wls}
             to obtain $\mathbf{h}_m^{n(t+1)}$

      \If{$\dfrac{\|\mathbf{h}_m^{n(t+1)}-\mathbf{h}_m^{n(t)}\|_2}
      {\|\mathbf{h}_m^{n(t)}\|_2+\epsilon}<10^{-8}$}
        \State \textbf{break}
      \EndIf
    \EndFor

    \State $\mathbf{h}_m^n\leftarrow\mathbf{h}_m^{n(T)}$
  \EndFor

  \State $\hat{\mathbf{h}}_m\leftarrow\mathbf{h}_m^N$
\EndFor

\State Form $\hat{\mathbf{H}}=[\hat{\mathbf{h}}_1,\ldots,
       \hat{\mathbf{h}}_M]^H$; compute $\mathbf{R}$
       via~\eqref{eq:cov_music}

\State Compute noise subspace $\mathbf{U}_N$; evaluate
       $P_Q(\theta)$ via~\eqref{eq:music_spectrum}

\end{algorithmic}
\end{algorithm}

\section{Numerical Results}
\label{sec:numerical}

\subsection{Simulation Setup}

We evaluate RobQMUSIC against Quantum-MUSIC of~\cite{kim2025} using
the same physical parameters.
Rydberg energy levels $52D_{5/2}$ and $53P_{3/2}$ of the  Caesium (Cs) atom detect RF signals at $\omega_{eg}\approx2\pi\times5$~GHz.
The transition dipole moment is obtained via the ARC
package~\cite{sibalic2017} as
$\boldsymbol{\mu}_{eg}=[0,1785.9\,qa_0,0]^T$, where
$q=1.602\times10^{-19}$~C and $a_0=5.292\times10^{-11}$~m.
Polarization directions $\boldsymbol{\epsilon}$ and
$\boldsymbol{\epsilon}_{LO}$ are i.i.d.\ $\mathcal{N}(0,\frac{1}{3})$,
and the LO reference power is $P_b=10 P_k$.
The pilot matrix $\mathbf{S}\in\mathbb{C}^{K\times P}$ has
i.i.d.\ $\mathcal{CN}(0,1)$ entries.
Throughout, we assume $K=2$ users, and they have true DoAs
$\theta_1=40^{\circ}$ and $\theta_2=-60^{\circ}$, and the angular
search grid contains $G_\theta=2000$ points on
$[-90^{\circ},90^{\circ}]$.

Three experiments are conducted and their specific parameters are
listed in Table~\ref{tab:sim_params}.
For the corruption sweep and spectrum comparison, $P_k$ and $\sigma_n^2$ are fixed at  ($P_k=10^{-18}$ and $\sigma_n^2=10^{-19.1}$.
For the SNR sweep, $P_k = 10^{-18}$ and
$SNR = \frac{P_k \boldsymbol{\mu_{eg}}(2) \frac{1}{3\hbar}}{\sigma_n^2}$ and we vary $\sigma_n^2$ to set the SNR value from 0 to 20 dB with an interval of 4dB. Outliers are injected by selecting a fraction $\eta$ of entries of $\mathbf{Z}$ uniformly at random and perturbing each by
$\pm\delta$, following~\eqref{eq:corrupted_meas} as done in \cite{Rajpurohit2025}, therefore making all the non-zero elements of $\mathbf{E}$ as $\pm \delta$.
The RMSE is
\begin{equation}
  \mathrm{RMSE}
  = \sqrt{\frac{1}{\mathrm{MC}\cdot K}
          \sum_{t=1}^{\mathrm{MC}}\sum_{k=1}^{K}
          \bigl(\hat\theta_k^{(t)}-\theta_k\bigr)^2}
  \label{eq:rmse_def}
\end{equation}
where $\mathrm{MC}$ denotes the number of Monte-Carlo runs.

\begin{table}[t]
  \centering
  \caption{Simulation Parameters}
  \label{tab:sim_params}
  \renewcommand{\arraystretch}{1.0}
  \setlength{\tabcolsep}{4pt} 
  \begin{tabular}{@{}lccc@{}}
    \toprule
    \textbf{Parameter} & \textbf{Fig. \ref{fig:qm_music_comparison}} & \textbf{Fig. \ref{fig:rmse_vs_corruption}} & \textbf{Fig. \ref{fig:rmse_vs_snr_combined}} \\
    \midrule
    Array size $M$             & 32        & 32        & 8             \\
    Snapshots $P$              & 100       & 500       & 200           \\
    Monte Carlo $MC$           & 1         & 100       & 500           \\
    Corruption $\eta$ (\%)     & 0,\;20    & 0--90     & 0,\;25        \\
    Outlier magnitude $\delta$ & 3         & 10        & 39            \\
    \midrule
    IRLS $\epsilon$            & \multicolumn{3}{c}{$10^{-8}$}        \\
    \bottomrule
  \end{tabular}
\end{table}
\subsection{MUSIC Spectrum Under Outlier Corruption}

\begin{figure}[!t]
\centering
\begin{subfigure}[t]{0.48\linewidth}
    \centering
    \includegraphics[width=\linewidth]{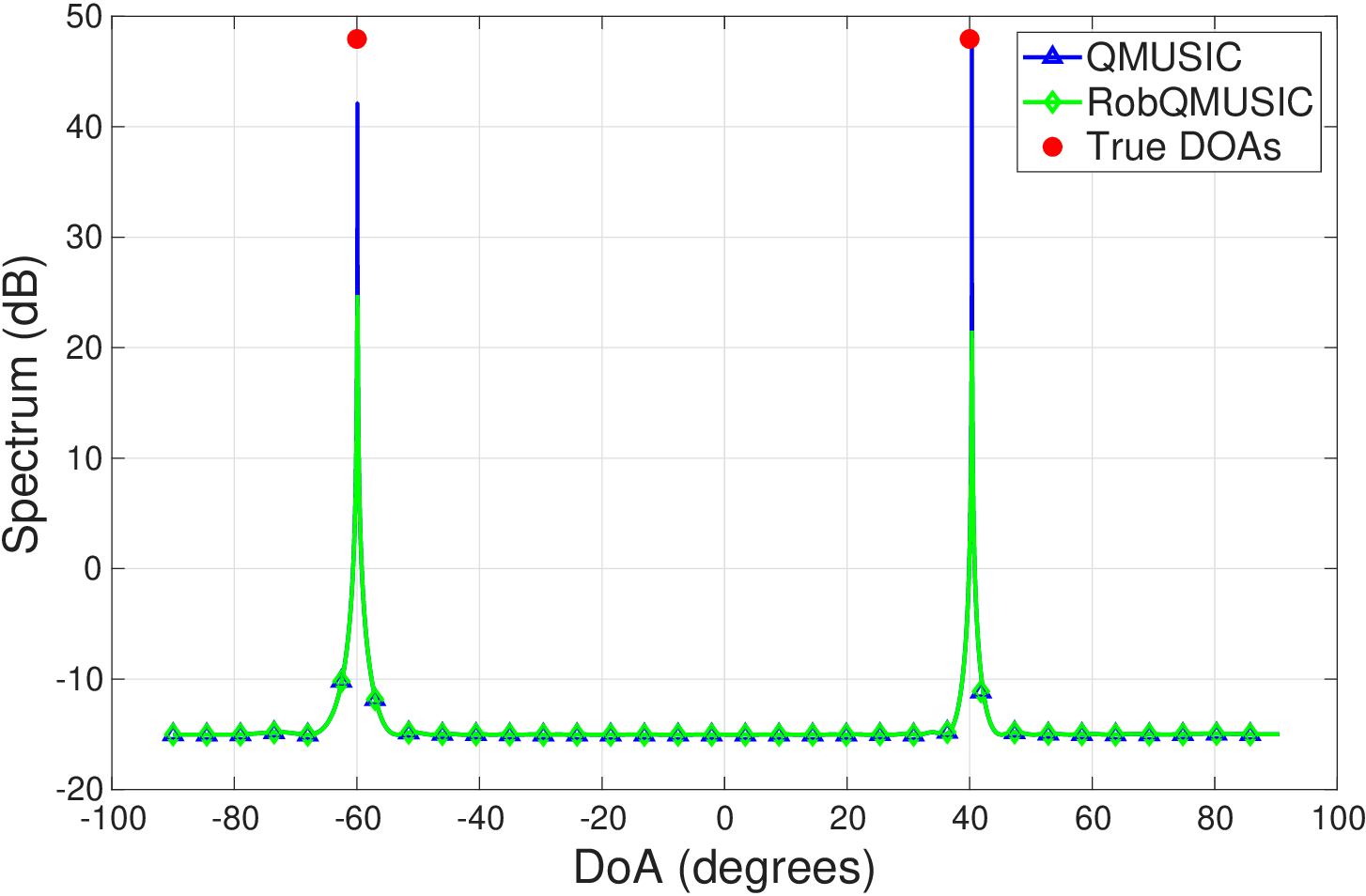}
    \caption{$\eta=0\%$}
    \label{fig:spectrum_clean}
\end{subfigure}
\hfill
\begin{subfigure}[t]{0.48\linewidth}
    \centering
    \includegraphics[width=\linewidth]{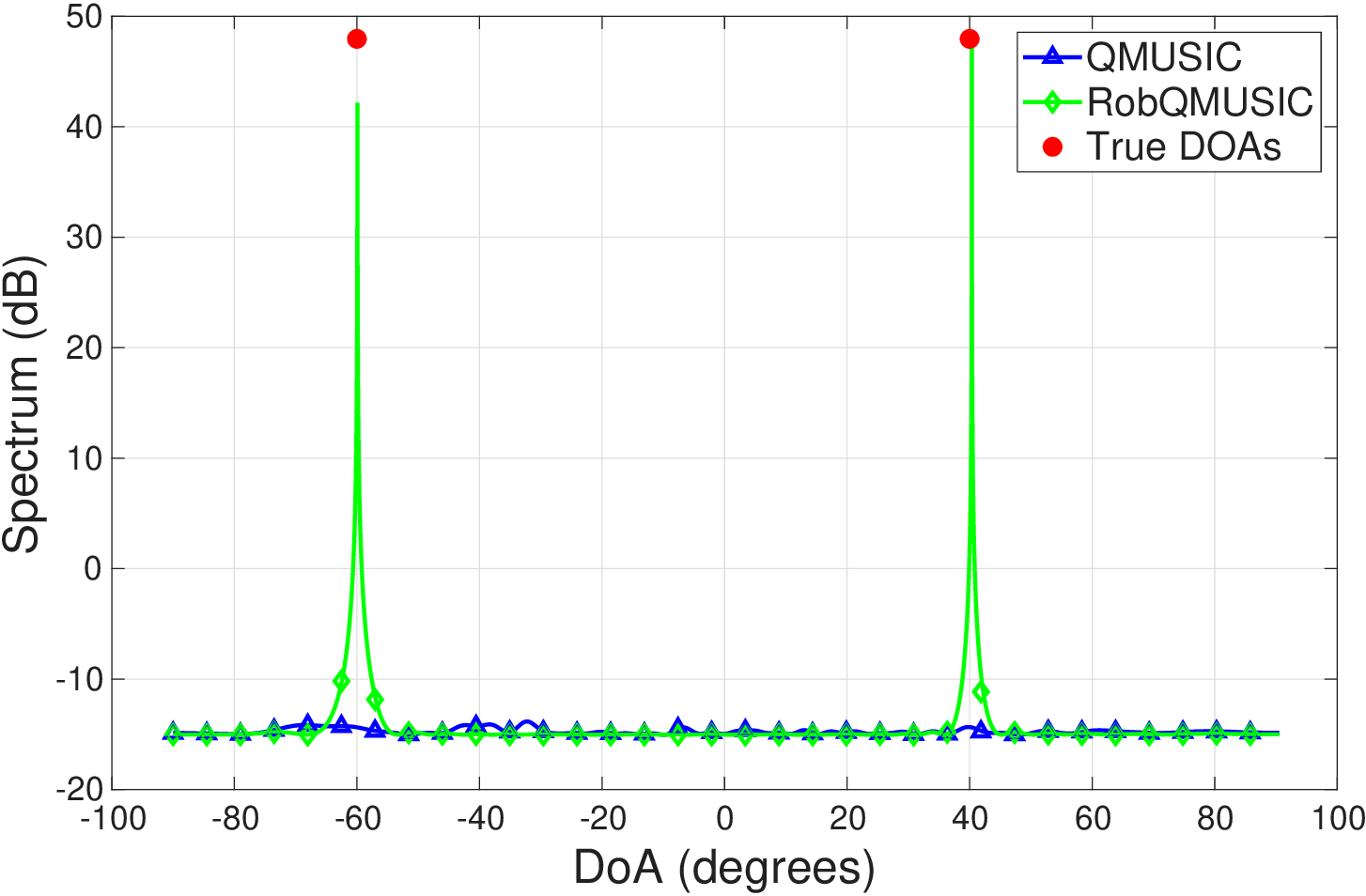}
    \caption{$\eta=20\%$}
    \label{fig:spectrum_corrupted}
\end{subfigure}

\caption{MUSIC and RobQMUSIC pseudo-spectrum for two different corruption levels.}

\label{fig:qm_music_comparison}

\end{figure}

Fig.~\ref{fig:qm_music_comparison} compares the MUSIC
pseudo-spectra for a single realisation at $\eta=0\%$ and
$\eta=20\%$. At $\eta=0\%$ (Fig.~\ref{fig:spectrum_clean}), both algorithms
produce sharp, well-separated peaks precisely aligned with the true
DoAs at $40^{\circ}$ and $-60^{\circ}$.
The spectral profiles are nearly indistinguishable, confirming that
the $\ell_1$-IRLS formulation does not incur a significant performance penalty in the absence of outliers.

At $\eta=20\%$ (Fig.~\ref{fig:spectrum_corrupted}), the contrast is
stark.
QMUSIC, which minimizes a quadratic residual, is overwhelmed by the
large-amplitude outliers: its spectrum is severely distorted and
fails entirely to resolve the two sources.
RobQMUSIC retains sharp, correctly located peaks at both DoAs,
demonstrating that the IRLS weight updates effectively isolate the
outlier-contaminated snapshots and preserve the signal-subspace
structure required for accurate MUSIC-based DoA estimation.

\subsection{RMSE Versus Outlier Fraction}

\begin{figure}[htbp]
  \centering
  \includegraphics[width=0.6\linewidth]{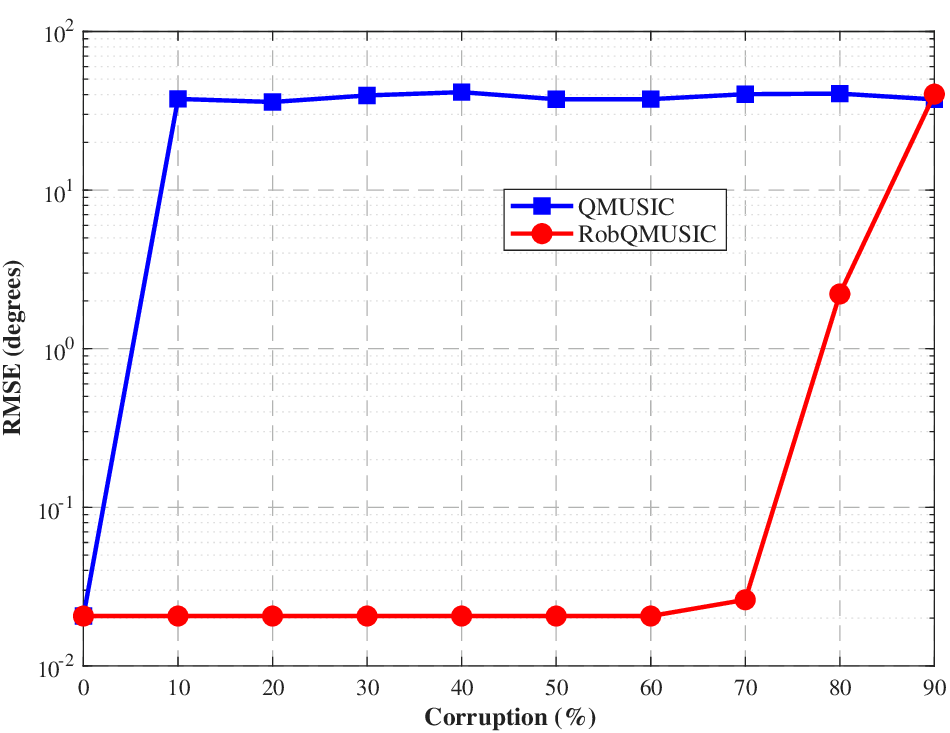}
  \caption{RMSE vs.\ outlier fraction $\eta$}
  \label{fig:rmse_vs_corruption}
\end{figure}

Fig.~\ref{fig:rmse_vs_corruption} shows the RMSE as a function of
the outlier fraction $\eta\in\{0,10,\ldots,90\}\%$ at
$\frac{P_k}{\sigma_n^2}\approx11$~dB. At $\eta=0\%$, both algorithms achieve RMSE $\approx0.02^{\circ}$,
confirming parity under ideal conditions.
As $\eta$ increases, QMUSIC degrades rapidly: the RMSE saturates near
$40^{\circ}$ for all $\eta\geq20\%$, indicating consistent failure
to estimate the DoA of the two sources.
RobQMUSIC maintains RMSE $\approx0.02^{\circ}$ across the entire
range $\eta\in[0\%,70\%]$. After $\eta = 70\%$, the algorithm fails to estimate the DoAs and loses its robustness. 

\subsection{RMSE Versus SNR}

\begin{figure}[!t]
\centering

\begin{subfigure}[t]{0.48\linewidth}
    \centering
    \includegraphics[width=\linewidth]{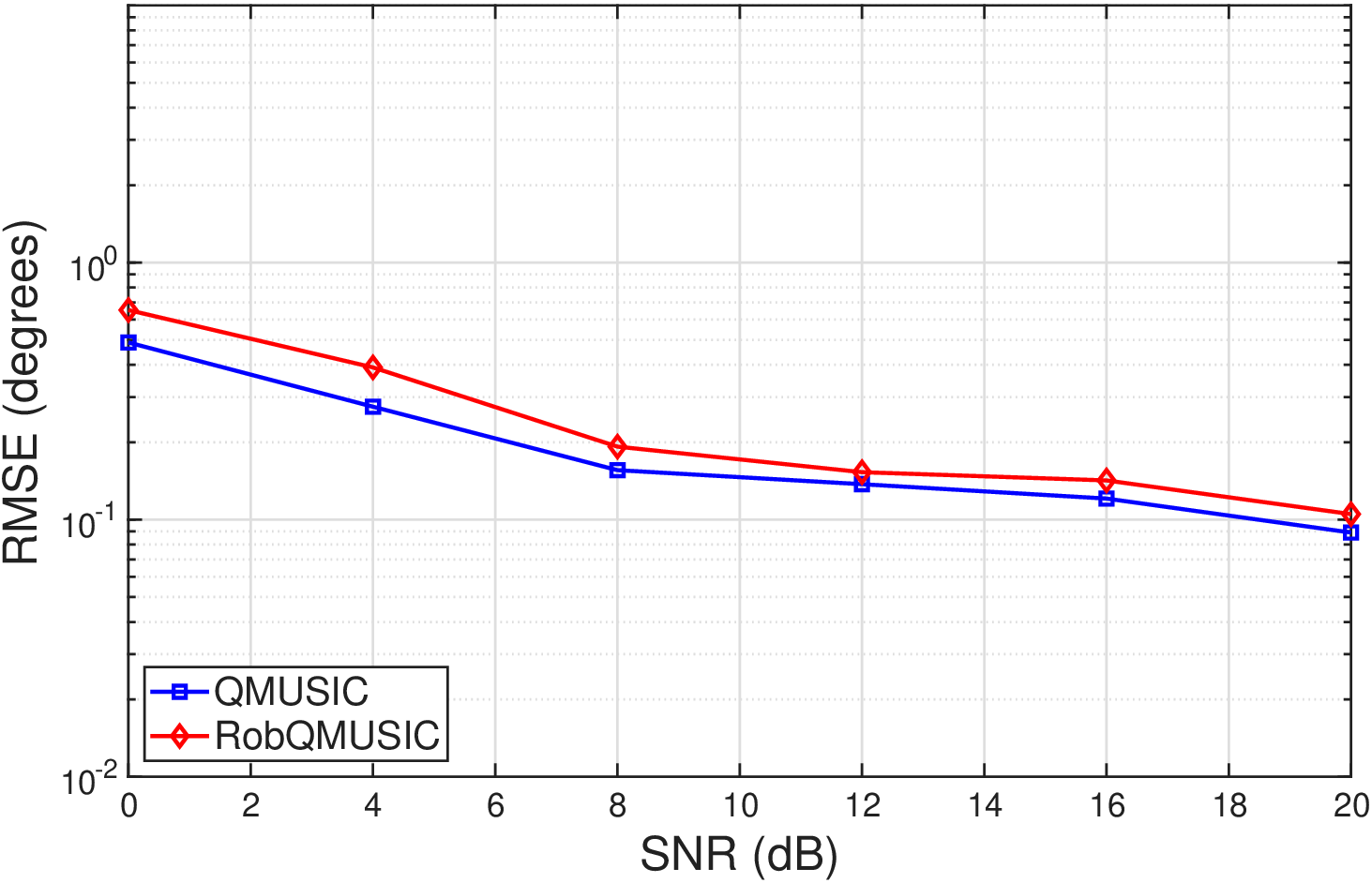}
    \caption{$\eta=0\%$}
    \label{fig:rmse_vs_snr_clean}
\end{subfigure}
\hfill
\begin{subfigure}[t]{0.48\linewidth}
    \centering
    \includegraphics[width=\linewidth]{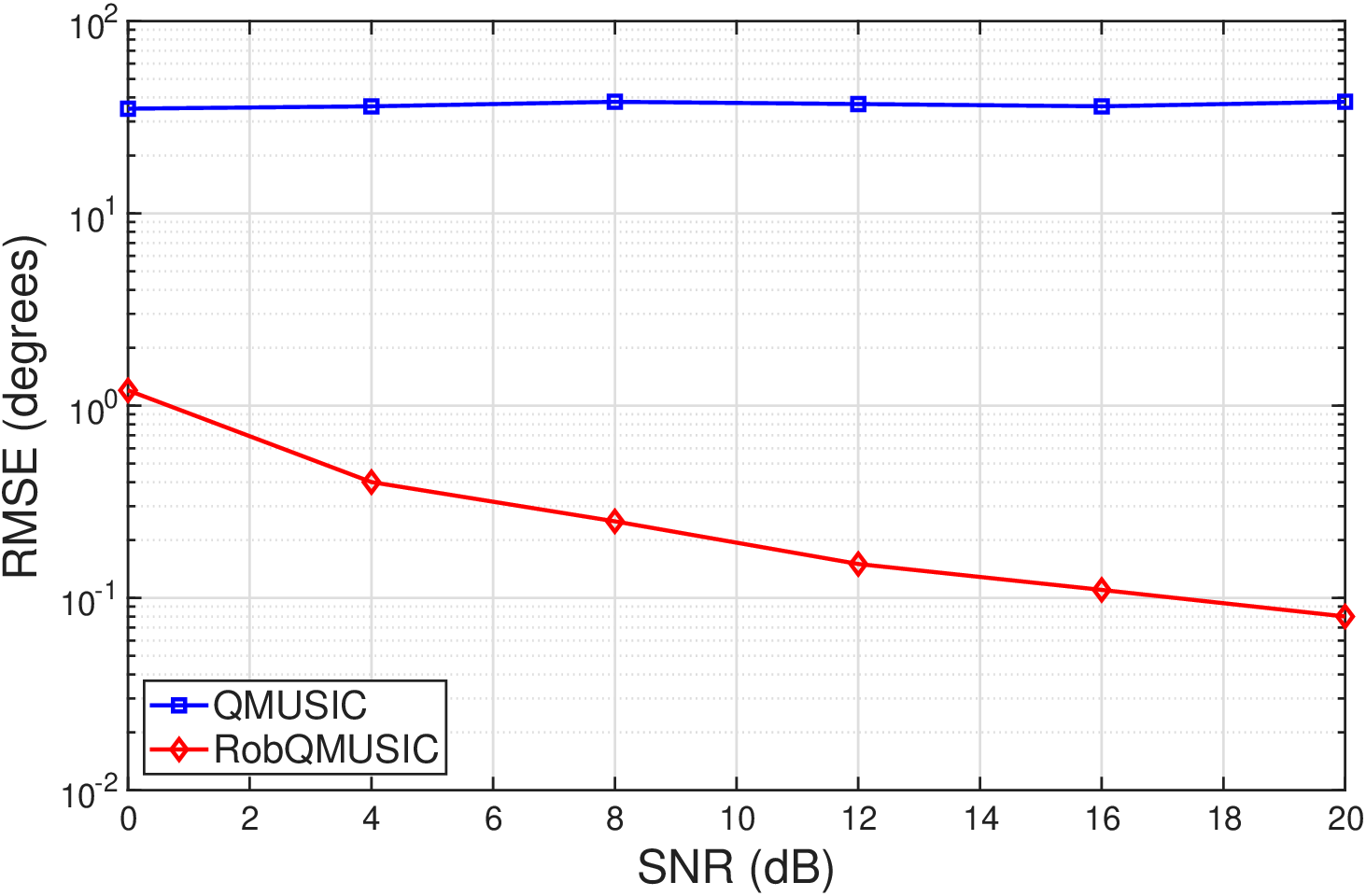}
    \caption{$\eta=25\%$}
    \label{fig:rmse_vs_snr_25}
\end{subfigure}

\caption{RMSE vs.\ SNR comparison under clean and corrupted measurements.}
\label{fig:rmse_vs_snr_combined}

\end{figure}

Figs. (\ref{fig:rmse_vs_snr_clean}) and (\ref{fig:rmse_vs_snr_25})
show the RMSE versus SNR at $\eta=0\%$ and $\eta=25\%$,
respectively ($M=8$, $P=200$, MC$=500$).
\subsubsection{No Outliers ($\eta=0\%$)}

Both algorithms improve monotonically with SNR.
QMUSIC achieves lower RMSE throughout which decreases with increasing SNR. 
RobQMUSIC follows the same trend with a slightly higher RMSE at each SNR value.
This gap reflects the inherent statistical efficiency cost of
$\ell_1$ penalisation: IRLS down-weights small-residual inliers
alongside outliers, raising the effective noise floor.
Nevertheless, both curves converge asymptotically at high SNR,
where the channel estimate is accurate enough that few snapshots
are misidentified as outliers.

\subsubsection{25\% Outlier Corruption ($\eta=25\%$)}

The situation reverses entirely under $25\%$ corruption.
QMUSIC saturates at $\approx35^{\circ}$ RMSE across the full SNR
range, confirming that the outlier energy completely dominates the
$\ell_2$ cost regardless of the operating SNR.
RobQMUSIC improves monotonically from $\approx1.2^{\circ}$ at
SNR $=0$~dB to $\approx0.07^{\circ}$ at SNR$=20$~dB.

Taken together, these experiments establish that RobQMUSIC
achieves a favorable robustness-efficiency trade-off. It incurs
a modest RMSE overhead relative to QMUSIC in the corruption-free
regime, in exchange for near-complete immunity to sparse
large-amplitude outliers—a trade-off that strongly benefits
practical quantum wireless sensing systems.


\section{Conclusion}
\label{sec:conclusion}

In this letter, we have proposed RobQMUSIC, a robust extension of the Quantum-MUSIC
algorithm for DoA estimation with Rydberg atomic receiver arrays.
By replacing the $\ell_2$-norm phase-retrieval step with an
$\ell_1$-IRLS formulation, the proposed algorithm effectively suppresses the
influence of sparse large-amplitude outliers without altering the
overall alternating-minimization structure or introducing
significant computational overhead.
Our numerical simulation results reveal that RobQMUSIC maintains near-identical
accuracy to Quantum-MUSIC under ideal conditions and achieves
reliable DoA estimation up to $80\%$ outlier contamination—a
regime in which Quantum-MUSIC degrades to near-worst-case error. Future works will focus on wideband quantum wireless sensing scenarios, as well as investigating advanced array geometries capable of estimating a larger number of sources than the number of physical elements under magnitude-only measurements.  


 \section*{Acknowledgment}
The authors would like to thank Prof. Prabhu Babu and Dr. Aakash Arora (CARE, IIT Delhi) for their valuable suggestions and insightful discussions that contributed to this work.


\bibliographystyle{IEEEtran}
\bibliography{refs}

@ARTICLE{kim2025,
  author={Kim, Hanvit and Park, Hyunwoo and Kim, Sunwoo},
  journal={IEEE Wireless Communications Letters}, 
  title={{Quantum-MUSIC: Multiple Signal Classification for Quantum Wireless Sensing}}, 
  year={2025},
  volume={14},
  number={6},
  pages={1623-1627},
  doi={10.1109/LWC.2025.3549767}}

@ARTICLE{candes2008introduction,
  author={Candes, Emmanuel J. and Wakin, Michael B.},
  journal={IEEE Signal Processing Magazine}, 
  title={{An Introduction To Compressive Sampling}}, 
  year={2008},
  volume={25},
  number={2},
  pages={21-30},
  doi={10.1109/MSP.2007.914731}}

@book{boyd2004convex,
place={Cambridge}, title={{Convex Optimization}},
publisher={Cambridge University Press}, 
author={Boyd, Stephen and Vandenberghe, Lieven}, 
year={2004}}

@misc{chen2026harnessingrydbergatomicreceivers,
      title={{Harnessing Rydberg Atomic Receivers: From Quantum Physics to Wireless Communications}}, 
      author={Yuanbin Chen and Xufeng Guo and Chau Yuen and Yufei Zhao and Yong Liang Guan and Chong Meng Samson See and Merouane Débbah and Lajos Hanzo},
      year={2026},
      eprint={2501.11842},
      archivePrefix={arXiv},
      primaryClass={cs.IT},
      url={https://arxiv.org/abs/2501.11842}, 
}

@ARTICLE{anderson2021,
  author={Anderson, David Alexander and Sapiro, Rachel Elizabeth and Raithel, Georg},
  journal={IEEE Transactions on Antennas and Propagation}, 
  title={{An Atomic Receiver for AM and FM Radio Communication}}, 
  year={2021},
  volume={69},
  number={5},
  pages={2455-2462},
  doi={10.1109/TAP.2020.2987112}}

@ARTICLE{fancher2021,
  author={Fancher, Charles T. and Scherer, David R. and John, Marc C. St. and Marlow, Bonnie L. Schmittberger},
  journal={IEEE Transactions on Quantum Engineering}, 
  title={{Rydberg Atom Electric Field Sensors for Communications and Sensing}}, 
  year={2021},
  volume={2},
  number={},
  pages={1-13},
  doi={10.1109/TQE.2021.3065227}}

@ARTICLE{robinson2021,
  author  = {Robinson, Amy K. and Prajapati, Nikunjkumar
             and Senic, Damir and Simons, Matthew T.
             and Holloway, Christopher L.},
  journal = {Applied Physics Letters},
  title   = {{Determining the Angle-of-Arrival of a
             Radio-Frequency Source with a {Rydberg}
             Atom-Based Sensor}},
  year    = {2021},
  volume  = {118},
  number  = {11},
  pages   = {114001},
  doi     = {10.1063/5.0045601}
}

@ARTICLE{mao2023,
  author={Mao, Ruiqi and Lin, Yi and Fu, Yunqi and Ma, Yuemin and Yang, Kai},
  journal={IEEE Transactions on Antennas and Propagation}, 
  title={{Digital Beamforming and Receiving Array Research Based on Rydberg Field Probes}}, 
  year={2024},
  volume={72},
  number={2},
  pages={2025-2029},
  doi={10.1109/TAP.2023.3327812}}

@ARTICLE{cui2025,
  author={Cui, Mingyao and Zeng, Qunsong and Huang, Kaibin},
  journal={IEEE Journal on Selected Areas in Communications}, 
  title={{Towards Atomic MIMO Receivers}}, 
  year={2025},
  volume={43},
  number={3},
  pages={659-673},
  doi={10.1109/JSAC.2025.3531528}}

@ARTICLE{schmidt1986,
  author={Schmidt, R.},
  journal={IEEE Transactions on Antennas and Propagation}, 
  title={Multiple emitter location and signal parameter estimation}, 
  year={1986},
  volume={34},
  number={3},
  pages={276-280},
  doi={10.1109/TAP.1986.1143830}}

@ARTICLE{candes2015,
  author={Candès, Emmanuel J. and Li, Xiaodong and Soltanolkotabi, Mahdi},
  journal={IEEE Transactions on Information Theory}, 
  title={{Phase Retrieval via Wirtinger Flow: Theory and Algorithms}}, 
  year={2015},
  volume={61},
  number={4},
  pages={1985-2007},
  doi={10.1109/TIT.2015.2399924}}

@ARTICLE{daubechies2010,
  author  = {Daubechies, Ingrid and DeVore, Ronald
             and Fornasier, Massimo
             and G{\"u}nt{\"u}rk, C. Sinan},
  journal = {{Communications on Pure and Applied Mathematics}},
  title   = {{Iteratively Reweighted Least Squares
             Minimization for Sparse Recovery}},
  year    = {2010},
  volume  = {63},
  number  = {1},
  pages   = {1--38},
  doi     = {10.1002/cpa.20303}
}

@ARTICLE{lai2013,
  author  = {Lai, Ming-Jun and Xu, Yangyang and Yin, Wotao},
  journal = {SIAM Journal on Numerical Analysis},
  title   = {{Improved Iteratively Reweighted Least Squares
             for Unconstrained Smoothed $\ell_q$ Minimization}},
  year    = {2013},
  volume  = {51},
  number  = {2},
  pages   = {927--957},
  doi     = {10.1137/110840364}
}

@ARTICLE{liu2023,
  author  = {Liu, Biao and Zhang, Li-Hua and Liu, Zhi-Kun
             and Zhang, Zhi-Yuan and Zhu, Zhi-Han
             and Gao, Wei and Guo, Guang-Can
             and Ding, Dong-Sheng and Shi, Bao-Sen},
  journal = {Electromagnetic Science},
  title   = {{Electric Field Measurement and Application
             Based on {Rydberg} Atoms}},
  year    = {2023},
  volume  = {1},
  number  = {1},
  pages   = {1--16},
  doi     = {10.23919/emsci.2022.0033}
}

@ARTICLE{sibalic2017,
  author  = {{\v{S}}ibali{\'c}, Nikola and Pritchard,
             Jonathan D. and Adams, Charles S.
             and Weatherill, Kevin J.},
  journal = {Computer Physics Communications},
  title   = {{{ARC}: an Open-Source Library for Calculating
             Properties of Alkali {Rydberg} Atoms}},
  year    = {2017},
  volume  = {220},
  pages   = {319--331},
  doi     = {10.1016/j.cpc.2017.06.015}
}

@INPROCEEDINGS{banerjee2026,
  author={Banerjee, Sourav and Kundu, Neel Kanth},
  booktitle={2026 IEEE Applied Sensing Conference (APSCON)}, 
  title={{Rydberg Atomic RF Sensor-based Quantum Radar}}, 
  year={2026},
  volume={},
  number={},
  pages={1-4},
  doi={10.1109/APSCON68325.2026.11497871}}

@ARTICLE{Rajpurohit2025,
  author={Rajpurohit, Pushpendra and Babu, Prabhu and Stoica, Petre},
  journal={IEEE Transactions on Aerospace and Electronic Systems}, 
  title={{Robust Direction-of-Arrival Estimation in the Presence of Outliers}}, 
  year={2025},
  volume={61},
  number={4},
  pages={10921-10927},
  doi={10.1109/TAES.2025.3560940}}

\end{document}